\documentclass[iop]{emulateapj}
\usepackage{graphicx,graphics}
\usepackage{longtable}
\usepackage{amsmath}
\usepackage{amssymb}

\newcommand{\whathow}[1]{\noindent\textbf{What:} #1\\ \textbf{How/Why :} \noindent\begin{list}{$\bullet$}{\itemsep -1.3ex \topsep -1.3ex}}
\newcommand{\ewhathow}{\end{list}}
\newcommand{\figref}[1]{Figure \ref{fig:#1}}
\newcommand{\tabref}[1]{Table \ref{tab:#1}}
\newcommand{\eqnref}[1]{Equation (\ref{eq:#1})}
\newcommand{\msun}{$M_\odot$}
\newcommand{\psr}{PSR\,J1614--2230}

\shorttitle{WD companion of a 2 Msun Neutron Star}
\begin{document}
\title{The White Dwarf Companion of a  2\,\msun\ Neutron Star}

\author{Varun B. Bhalerao and S. R. Kulkarni}
\affil{Department of Astronomy, California Institute of Technology, Pasadena, CA 91125, USA}

\begin{abstract}
We report the optical discovery of the companion to the 2\,\msun\
millisecond pulsar \psr. The optical colors show that the 0.5\,\msun\
companion is a 2.2\,Gyr old He--CO white dwarf. We infer that $\dot{M}$ during the accretion phase is $<10^{-2}\,\dot{M}_{\rm edd}$. We show that the pulsar was born with a spin close to its current value, 
well below the rebirth line. The spin-down parameters, the mass of the pulsar, and the age of the system challenge the simple recycling model for the formation of millisecond pulsars.
\end{abstract}


\maketitle

\section{\psr}

\psr, a 3.15-\,ms pulsar,
was discovered in a radio survey of unidentified EGRET
gamma ray sources using the Parkes Radio Telescope \citep{discovery}.
Subsequently, X-ray emission  from {\em Newton XMM} \citep{xraypwn}
and $\gamma$-ray emission from {\em Fermi Gamma Ray Large Area Space Telescope} \citep{fermi} was detected.
Like most millisecond pulsars (MSPs), \psr\ 
is in a binary system. The circular orbit is consistent with the pulsar 
having undergone mass transfer and spun up. The mass function derived
from pulsar timing indicated a companion with mass $M_2>0.4\,M_\odot$
\citep{discovery}.

The system recently came into prominence when \citet{heavy} reported
the mass of the pulsar to be $1.97\pm0.04\,$\msun. The detection
of such a massive neutron star (NS) places very strong constraints on
the equation of state of matter at extreme nuclear densities
\citep[see, for example,][]{lattimer,lattimertable}.  The rather exquisite precision of
this mass measurement was possible due to the orbit being 
almost perpendicular to the plane of the sky. As a result,
the Shapiro delay caused by the companion is very large, resulting in
a precise estimate of the mass of the companion, $M_2=
0.500\pm 0.006\,M_\odot$. The 8.7\,day orbital period is significantly shorter than $\sim120\,$days expected for a low-mass X-ray binary with such a massive secondary \citep{massperiod} --- suggesting a peculiar evolutionary history for this binary.

Given the importance of the result of \cite{heavy} additional verification
or consistency checks of physical parameters of \psr\
can be expected to be of some value. 
A  0.5\,\msun\ white dwarf (WD) at the inferred distance of \psr\ ($d\sim 1.2$\,kpc),
even if a few Gyr old, is within the reach
of present-day optical telescopes. It is this search for the WD that constitutes the
principal focus of this Letter.

\section{Observations at the W. M. Keck Observatory}\label{sec:observations}

We observed \psr\  (\figref{detect}) in $g$ and $R$
bands using the imaging mode of the Low Resolution Imaging Spectrograph (LRIS)
on the 10\,m Keck-I telescope~\citep{lris}, with upgraded
red\footnote{http://www2.keck.hawaii.edu/inst/lris/lris-red-upgrade-notes.html}
and blue cameras~\citep{lrisb1,lrisb2}.  Several images were acquired
at each target location, dithering the telescope by small amounts
between each exposure. The observing conditions on UT
2010 May 15 were poor (seeing 1\arcsec.4), so only data acquired on UT 2010 July 8 ($R$ band seeing 0\arcsec.85 FWHM) were used in
this analysis. The total exposure on this night was 960\,s in the $R$ band and 1010\,s in the $g$ band. The plate scale is 0\arcsec.135\,pixel$^{-1}$ for both cameras

\begin{deluxetable}{lllll}[!hbtp]
\tablecaption{Positions and Magnitudes of Reference Stars\label{tab:refs}}
\tablewidth{0pt}
\tablehead{
\colhead{ID} & \colhead{RA} & \colhead{Declination} & \colhead{$m_R$\tablenotemark{a}} & \colhead{$m_g$} \\
& \colhead{$\alpha - 16^{\rm h}14^{\rm m}$} & \colhead{$\delta - (-22\degr)$} & 
}
\startdata
A & 36$^{\rm s}$.98(8) & 29\arcmin 18\arcsec.78(11) & 17.94(8)   & 18.73(3)  \\
B & 36$^{\rm s}$.29(7) & 29\arcmin 31\arcsec.97(7) & 17.71(8)   & 18.38(3)  \\
C & 35$^{\rm s}$.88(9) & 30\arcmin 18\arcsec.70(10) & 19.76(10)   & 20.41(4)  \\
D & 34$^{\rm s}$.39(5) & 30\arcmin 12\arcsec.11(12) & 19.15(8)   & 21.24(4) \\
E\tablenotemark{b} & 35$^{\rm s}$.66 & 31\arcmin 04\arcsec.17  & 21.38(9)   & 22.04(5) \\
F & 34$^{\rm s}$.92(16) & 30\arcmin 59\arcsec.46(2) & 20.18(9)   & 21.36(5) \\
G\tablenotemark{b} & 37$^{\rm s}$.50 & 30\arcmin 43\arcsec.69  & 20.71(8)   & 21.87(4) \\
\tableline
Q & 36$^{\rm s}$.47(7) & 30\arcmin 35\arcsec.90(4) & (Saturated) & 17.20(3)  \\
R & 35$^{\rm s}$.92(2) & 30\arcmin 30\arcsec.45(1) & 20.10(8) & 20.94(4) \\
S & 36$^{\rm s}$.50(10) & 30\arcmin 13\arcsec.93(13) & (Saturated) & 17.75(3) 
\enddata
\tablecomments{Stars A--G were used as reference stars for photometry. Right ascension and declination were obtained from USNO-B1.0 unless otherwise specified.}
\tablenotetext{a}{$R$ band magnitudes calculated using SDSS magnitudes and \citet{sdssequations} transformation equations (Section~\ref{sec:observations}). The numbers in parenthesis do not include a 0.03\,mag uncertainty in absolute calibration due to the transformations.}
\tablenotetext{b}{Coordinates obtained from our final $R$-band science image. The World Coordinate System for this image was calculated using a total of 33 USNO-B1.0 stars (Section~\ref{sec:observations}).}
\end{deluxetable}

\begin{deluxetable*}{rlllll}
\tablecaption{Coordinates of \psr\ at Different Epochs}
\tablewidth{6.5in}
\tablehead{
\colhead{Method} & \colhead{Epoch} & \colhead{Ecliptic} & \colhead{Ecliptic} & \colhead{RA}& \colhead{Declination} \\
 & & \colhead{Longitude ($\lambda$)} & \colhead{Latitude ($\beta$)} & \colhead{$\alpha\ - $ 16$^{\rm h}$14$^{\rm m}$} & \colhead{$\delta\ - $ ($-22$\degr30\arcmin)}
}
\startdata
Timing & J2005.63 & 245.78827556(5) & $-$1.256744(2) & 36$^{\rm s}$.5051(1) &  31\arcsec.080(7) \\
Timing & J2000.00 & 245.78826025(12) & $-$1.256697(5) & 36$^{\rm s}$.5034(2) &  30\arcsec.904(19) \\
\tableline
Timing & J2007.32 & 245.78828015(6) & $-1$.256758(2) & 36$^{\rm s}$.5056(1) & 31\arcsec.132(9) \\
{\em Chandra}\tablenotemark{a} & J2007.32 & \nodata & \nodata & 36$^{\rm s}$.50(15) & 31\arcsec.33(20) \\
\tableline
Timing & J2010.51 & 245.78828886(11) & $-$1.256785(5) & 36$^{\rm s}$.5067(2) & 31\arcsec.23(2) \\
LRIS & J2010.51 & \nodata & \nodata & 36$^{\rm s}$.50(16) & 31\arcsec.72(20) \\
\enddata
\tablecomments{
The proper motion of the pulsar as obtained
from timing observations is $\mu_\lambda=9.79(7)$\,mas\,yr$^{-1}$, $\mu_\beta = -30(3)$\,mas\,yr$^{-1}$. This proper motion is used to estimate the ecliptic coordinates of the pulsar at the epoch of the {\em Chandra} and LRIS observations. The right ascension and declination are calculated from ecliptic coordinates using the \texttt{Euler} program in \texttt{IDL}. The equinox in all cases is J2000.
}
\tablenotetext{a}{Bore-sight corrected coordinates. The target and the reference source ``E'' are detected in the {\em Chandra} image. We extract source coordinates using \texttt{celldetect}. We assume that the {\em Chandra} coordinate system and our $R$-band coordinate solution are related by a simple offset with no rotation. Using the X-ray and $R$-band coordinates of source E, we calculate that the offset is $\alpha_X - \alpha_R = 0\arcsec.21$, $\delta_X - \delta_R = -0\arcsec.20$ and correct the target position using this offset.}
\label{tab:RADEC}
\end{deluxetable*}

The images were processed using \texttt{IRAF}. After bias correction and flat fielding, cosmic rays were rejected using \texttt{L.A.Cosmic}~\citep{lacosmic}. The images were then aligned with \texttt{xregister} and averaged to produce the final image for each band
(see \figref{detect}).
 A World Coordinate System was calculated using USNO-B1.0 stars in the field, with the \texttt{imcoords} package. Using 33 stars in the field, for the final $R$-band image we obtained an rms error of 0\arcsec.14 in right ascension (R.A.) and 0\arcsec.20 in declination, adding up to a radial error of 0\arcsec.24. For the final $g$-band image, the residuals were 0\arcsec.13 and 0\arcsec.18 for R.A. and declination, respectively, giving a total error of 0\arcsec.22.

We measured fluxes with aperture photometry using the \texttt{IDL APPHOT} package. For each night, we measured the seeing (FWHM) and set the aperture to one seeing radius~\citep{aperturephot}. We extracted the sky from an annulus 5--10 seeing radii wide. We had observed nearby a Sloan Digital Sky Survey \citep[SDSS;][]{thesdss} field\footnote{Calibration SDSS field: $\alpha = 17^{\rm h}$19$^{\rm m}$10$^{\rm s}$.10, $\delta=
  -14\degr$38\arcmin46\arcsec.0.} with the same settings as the science field. The calibration field was observed immediately after the science exposures, and had airmass 1.4 as compared to 1.6 for the target.  We used the magnitudes of six stars from that field to calculate the photometric zero point and calibrated six reference stars in the science field (\figref{detect}, \tabref{refs}). $R$ band magnitudes were calculated using photometric transformations\footnote{Photometric transforms: $V = g - 0.59(g-r) - 0.01 \pm 0.01$ and $V-R = 1.09(r-i) + 0.22 \pm 0.03$.} prescribed by \citet{sdssequations} for stars with $R_c - I_c < 1.15$. The typical uncertainty in $g$-band magnitudes for stars with $m_g \sim 20$ is $0.03$\,mag. In $R$ band we have $\Delta m_R = 0.07$ for $m_R \sim 20$, including the uncertainty in the transformations.

\section{Detection of an Optical Counterpart}

\begin{figure*}[htbp]
  \centering
  \includegraphics[width=4.5in]{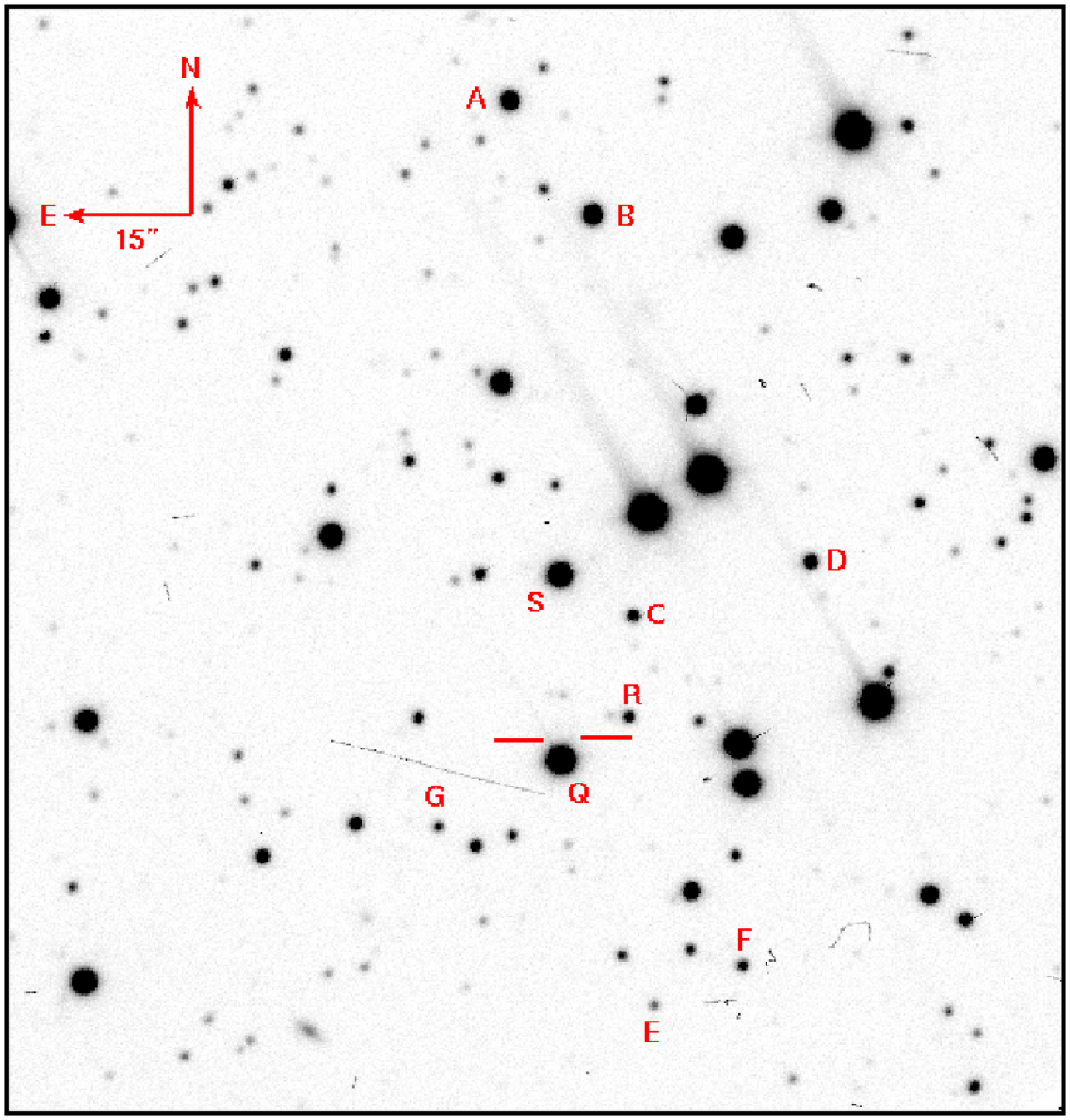}
  \includegraphics[width=4.5in]{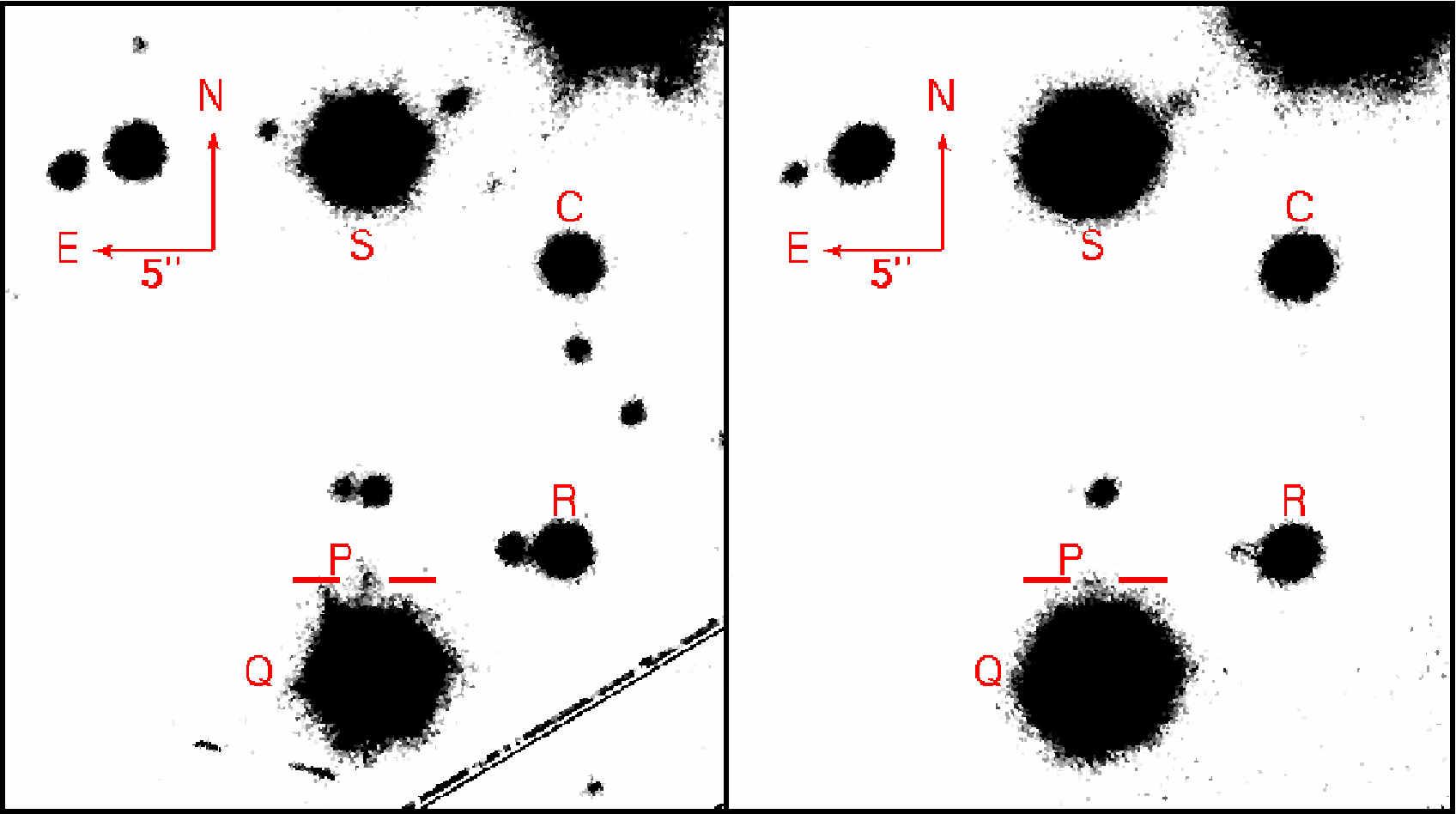} 
  \caption{Top: $R$ band LRIS image of the \psr\ field. The expected location of the target for epoch 2010.52 is shown by two horizontal lines just below the center of the image. The CCD shows considerable
  blooming in rows and columns around bright stars. The hexagonal
  mirror shape gives six diffraction spikes separated by 60\degr. To
  avoid contamination of the target by these artifacts from the
  bright star Q, we set the position angle to 300\degr.
  Bottom: $R$-band (left) and $g$-band (right) images of \psr. The target (P) is detected in the $R$
   band at $\alpha = 16^{\rm h}$14$^{\rm m}$36$^{\rm s}$.50, $\delta=
  -22\degr$30\arcmin31\arcsec.72 and is marked with thick horizontal lines.
  There is a 0\arcsec.33 offset between the expected and observed positions
  of the target (\tabref{RADEC}). The target is not detected in the $g$-band, but the $R$ band location is marked for reference.
  The diagonal streak in the $R$ image is a bad CCD column.}
  \label{fig:detect}
\end{figure*}

In the vicinity of the nominal pulsar position, we find a faint
source (labeled ``P'') in the $R$ band (\figref{detect}). We do not detect anything
within 1\arcsec\ of the target in the $g$ band. 
The optical coordinates of this source, the timing position of the pulsar and
a proposed X-ray counterpart are summarized in 
\tabref{RADEC}. To compare this with the source location, we first have to correct for the 33\,mas\,yr$^{-1}$ proper motion of the source.
The LRIS
source P is about 0\arcsec.50 South of the pulsar position (extrapolated
for the epoch of LRIS observations). Given the 0\arcsec.24
(1-$\sigma$) astrometric uncertainty of the optical images, this position is consistent
with the location of the pulsar.

The density of objects brighter than star P in this field is 0.02\,arcsecond$^{-2}$. Using the 0\arcsec.85 seeing FWHM as the mean diameter of stars, we calculate a false identification probability of $1\%$. The excellent astrometric coincidence and the low probability of chance coincidence embolden us to suggest that star P is the optical counterpart of \psr.

Counterpart P is located only 4\arcsec.2 from the 16.3 mag
star \object[USNO-B1.0 0674-0429635]{USNO-B1.0 0674-0429635}, and 
is contaminated by the flux in the wings of its point-spread function.
The proximity to this bright source will bias both the photometry and
the astrometry of the counterpart. To calculate the 
bias, we injected fake Gaussian sources with FWHM matched to the
seeing and brightness comparable to the faint object.  
We then measured the coordinates and
magnitude of the injected source using the same procedure as for
the faint object. We find that for separations $\approx 4$\arcsec.2,
the recovered position is systematically pulled towards the 
bright star by 0\arcsec.1 -- 0\arcsec.2. This is small enough that we do not apply this correction. The same exercise led us to derive the photometric bias. The de-biased $R$-band magnitude of P is $m_R = 24.3 \pm 0.1$.

\begin{figure*}[htbp]
  \centering
  \includegraphics[scale=0.8,angle=270]{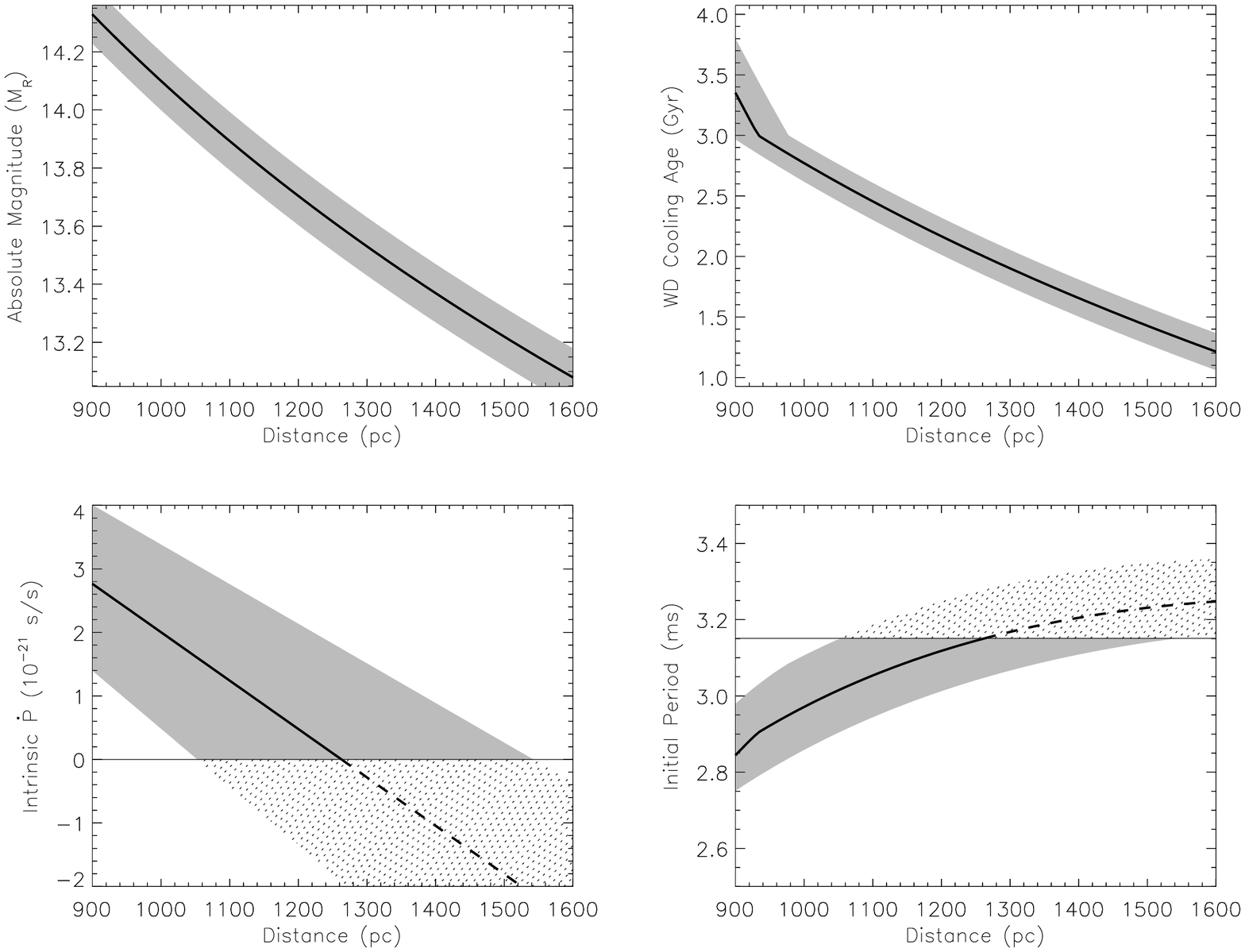} 
  \caption{Inferred parameters for \psr\ as a function of distance. The gray or dotted regions show the error bars on each parameter. Top left: The absolute $R$-band magnitude of the WD. Top right: WD cooling age inferred from \citet{wdcooling}. The kink at 3\,Gyr may be a result of granularity of the tables. Bottom left: the intrinsic period derivative ($\dot{P}$) in the pulsar frame, corrected for the Shklovskii effect. Since the pulsar cannot be spinning up, the values of $\dot{P} < 0$ are unphysical and are shown as the dotted area. The maximum distance to the pulsar is inferred to be 1540\,pc. Lower right: the initial spin period of the pulsar. Birth periods slower than the current period are excluded.}
  \label{fig:dist}
\end{figure*}

To infer properties of the WD, we need to calculate its absolute
magnitude. 
To calculate the optical extinction, we assume that the ratio $N_{\rm H}/N_e$ is constant along the line of sight in the direction of \psr. The dispersion measure (DM) for \psr\ is 34.5\,pc\,cm$^{-3}$~\citep{heavy}, and the total DM in this direction is 104\,pc\,cm$^{-3}$~\citep{dispersion} --- about a factor of three higher than the pulsar DM. This implies that the pulsar is behind approximately one third of the galactic absorbing column. We can then scale the total $R$ band extinction in this direction \citep[$A_{R,{\rm total}} = 0.65$;][]{extinction} to get $A_R = 0.22$. We assume $\lambda_{\rm eff} = 0.47\,\mu$m for the $g$ band\footnote{http://www.sdss.org/dr6/instruments/imager/index.html\#filters}, and use the standard reddening law with $R_V = 3.1$ to get $A_g = 0.34$~\citep{allens}.

\citet{heavy} use the DM to estimate that \psr\ is at a distance $d = 1.2$\,kpc. For this distance, the extinction-corrected $R$-band absolute magnitude is $M_R = 13.7\pm0.1$. Furthermore, they place a lower limit of 900\,pc on the distance, using pulsar parallax from timing measurements. The corresponding absolute magnitude is $M_R = 14.3\pm0.1$.

\section{Pulsar age and birth spin period}\label{sec:age}
%

The measured mass (0.5\,\msun) and the inferred absolute magnitude ($M_R \approx 13.7$) when applied to WD cooling models~\citep{wdcooling} lead to an age of $\tau_{\rm WD} \sim 2.2$\,Gyr. For such a WD, the expected absolute magnitude in other bands is $M_B = 14.6$ and $M_V = 14.1$. This gives $m_g \approx 25.0$, consistent with our non detection.

As per the standard evolutionary model for MSPs, the NS in \psr\ was spun up by accretion from a low- or intermediate-mass companion star. The accretion stopped when the companion decoupled from its Roche lobe and became a WD. At this point, the MSP started spinning down by radiating energy. The spin-down age of the MSP is thus equal to the cooling down age of the WD. 

The period of the pulsar at birth ($P_{\rm init}$), its spin-down age ($\tau$), present-day period ($P$) and period derivative ($\dot{P}$) are related to each other as follows: 

\begin{equation}\label{eq:age}
 \tau = \frac{P}{(n-1)\dot{P}} \left[ 1 - \left( \frac{P_{\rm init}}{P} \right)^{n-1} \right]
\end{equation}

\noindent where $n$ is the ``braking index'' for the pulsar, with $n=3$ appropriate for a dipole radiating into vacuum. Thus, the period at birth is given by 
\begin{equation}\label{eq:pinit}
P_{\rm init} = P \left[ 1 - \frac{\tau \dot{P} (n-1) }{P} \right]^{1/(n-1)}
\end{equation}


The measured period derivative of a pulsar ($\dot{P}_{\rm obs}$) is always higher than its true period derivative ($\dot{P}$) owing to transverse motion~\citep{shklovskii}. The corrected period derivative is $\dot{P} = \dot{P}_{\rm obs} - P\mu^2 d/c$,
where $d$ is the distance and $\mu$ is the proper motion. \citet{heavy} measure $\mu = 32(3){\rm\,mas\,yr}^{-1}$ for \psr. Using the nominal distance $d=1.2\,$kpc, $\dot{P} = 4.8\times 10^{-22}{\rm\,s\,s}^{-1}$.

\begin{figure}[htbp]
  \centering
  \includegraphics[scale=0.65, viewport=30 0 432 432]{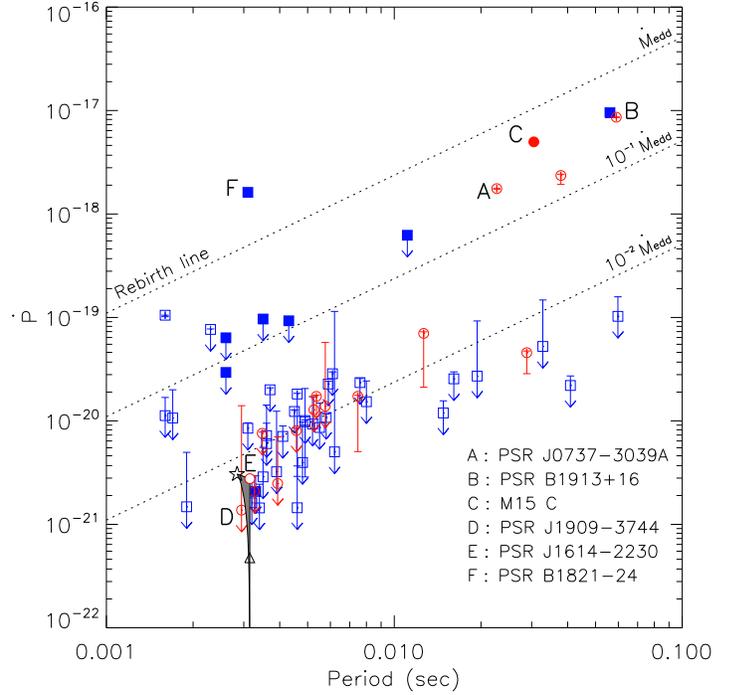} 
  \caption{Pulsar $P$--$\dot{P}$ diagram. Period derivatives are corrected for the Shklovskii effect. Some interesting systems are labeled. Filled symbols are pulsars in globular clusters, whose $\dot{P}$ may be affected by the dense environment. Circles denote NS binaries with mass measurements. The dashed lines are the pulsar rebirth lines $\dot{P} = (\dot{M}/\dot{M}_{\rm edd}) 1.1\times10^{-15}{\rm\,s}^{-4/3} P^{4/3} $~\citep{peq} for $\dot{M}/\dot{M}_{\rm edd} = 1$, $10^{-1}$ and $10^{-2}$. The location of \psr\ on this plot depends on the distance (see Section~\ref{sec:age}). Were the pulsar to be at 900\,pc, it would be born at the star symbol and evolve toward its present-day location, the hollow circle marked ``E''. If instead it is at 1200\,pc, the birth location is shown by the upward triangle, which also coincides with its present-day location. The solid curve passing through the star and the upward triangle denotes all possible birth locations for \psr. The system evolves through the shaded gray area to its present-day location on the vertical line through E and the upward triangle.}
  \label{fig:ppdot}
\end{figure}

The DM inferred distance is quite uncertain, so it is useful to consider the dependence of all parameters on distance. \figref{dist} shows the range of values for the $R$-band absolute magnitude ($M_R$), the WD cooling age ($\tau_{\rm WD}$), the intrinsic period derivative in the pulsar's frame ($\dot{P}$) and the initial spin period ($P_{\rm init}$). Since there is no energy injection to the pulsar now, it must be currently spinning down. This implies $\dot{P} \geq 0$ and allows us to calculate an upper limit on the distance to the pulsar: $d_{\rm max} = 1540\,$pc (1-$\sigma$). The last panel in \figref{dist} shows $P_{\rm init} \geq 2.75$\,ms. We conclude that the pulsar in \psr\ must have been born with a period close to the current observed value.

The birth spin period of a NS is governed by an equilibrium between the ram pressure of the accreting material and the magnetic field. NSs spun up by accretion at the Eddington rate ($\dot{M}_{\rm edd}$) are reborn as MSPs on the ``rebirth line''~\citep{peq}. \figref{ppdot} shows this rebirth line on a $P$--$\dot{P}$ diagram, along with current positions of various pulsars from the ATNF database~\citep{atnf}. Also shown are rebirth lines for pulsars accreting at $10^{-1} \dot{M}_{\rm edd}$ and $10^{-2} \dot{M}_{\rm edd}$. For a pulsar radiating as a dipole, the $P\dot{P}$ product remains constant through its lifetime. Thus, we can calculate the initial spin-down rate of the pulsar: $\dot{P}_{\rm init} = P \dot{P}/P_{\rm init}$, where $P_{\rm init}$ comes from \eqnref{pinit}. If \psr\ is at 900\,pc, then it would have been born with $\dot{P}_{\rm init} = 3.1\times 10^{-21}{\rm\,s\,s}^{-1}$ and $P_{\rm init} = 2.84\,$ms. This value is indicated with a star. The pulsar then evolves toward the lower right, to the circle marked ``E''. Similarly, the birth parameters for the pulsar assuming $d=1.2\,$kpc are shown by the upward triangle --- it is nearly coincident with the current parameters of the pulsar for this distance. Other possible birth locations for the pulsar lie along the solid line passing through the star and triangle. The gray region shows all possible positions that \psr\ can have occupied in its lifetime.  It is clear that the pulsar was born well below the rebirth line, and the mean accretion rate during the spin-up phase (the final major accretion phase) was lower than $10^{-2} \dot{M}_{\rm edd}$.

Two groups have run detailed simulations of the evolution of \psr. The \citet{evolve1} model and the preferred model of \citet{evolve2} are qualitatively similar: the system begins as an intermediate mass X-ray binary consisting of a NS and a $\sim 4\,$\msun\ main-sequence secondary, which evolves to form a CO WD with an He envelope. The secondary accretes mass onto the NS in three phases. The first phase (A1) is a thermal timescale mass transfer at super-Eddington accretion rates, where the NS gains little mass. The next phase (A2) is on a nuclear timescale ($\sim 35\,$Myr), when the secondary is burning H in the core and envelope. During this phase, the accretion rates are upto about a tenth of the Eddington limit. During the final accretion phase (phase AB), the secondary is burning He in its core and H in an envelope. This causes the radius of the donor star to expand, triggering accretion at near-Eddington rates for 5 -- 10\,Myr. The NS gains the most mass during this phase. For typical NS parameters, accreting 0.1 -- 0.2\,\msun\ is enough to spin them up millisecond periods~\citep{newage}. Hence, the near-Eddington accretion in phase AB should spin the pulsar up all the way to the rebirth line. This is inconsistent with our inferred birth position of \psr.

If the stellar evolution models are correct, then there is a problem with the standard formation scenario of MSPs~\citep{recycling1,recycling2}.
In essence, the birth period depends on factors other than the magnetic field strength and accretion rates. For instance, \citet{lars1,lars2} proposes that accretion induces a quadrupole moment $Q$ in the NS. The NS then loses angular momentum by gravitational wave radiation. Since the magnetic fields do not play any significant role in this model, the rebirth line becomes irrelevant, and \psr\ may start its life anywhere on the $P$--$\dot{P}$ diagram.
In summary, the birth of \psr\ away from the rebirth line requires reconsideration of angular momentum loss mechanisms.

Going forward, on-going radio observations (pulsar timing and VLBI) should improve the parallax and thereby
decrease the uncertainty in the inferred $\dot P$ and thus better locate the pulsar in the $P$-$\dot P$
diagram. The scattered light from light Q, while bothersome to the present observations, provide an opportunity
to use adaptive optics to measure the near-IR fluxes of the WD. Grism spectroscopy with {\em Hubble Space Telescope}
can potentially reveal the presumed H+He layer posited by stellar evolutionary models of \citet{evolve2}.

\noindent{\it Acknowledgments.}
We are grateful to Scott Ransom for providing the source position in advance of the publication. We thank M. Kasliwal and A. Gal-Yam for undertaking the observations.

Some of the data presented herein were obtained at the W.M. Keck Observatory, which is operated as a scientific partnership among the California Institute of Technology, the University of California and the National Aeronautics and Space Administration. The Observatory was made possible by the generous financial support of the W.M. Keck Foundation.


{\it Facilities:} \facility{Keck:I (LRIS)}

\end{document}